\documentclass[11pt]{article}
\usepackage{amsmath,amssymb,amsfonts}
\usepackage{latexsym}
\usepackage{pstricks}  

\begin{document}
\begin{titlepage}
\begin{flushright}
UTAS-PHYS-2009-28\\
December 2009\\
\end{flushright}
\begin{centering}
 
{\ }\vspace{0.5cm}
 
{\Large\bf Exactly solvable three-level quantum dissipative systems}

\vspace{5pt}

{\Large\bf  via bosonisation of fermion gas-impurity models }

\vspace{1.8cm}

Sol H. Jacobsen\footnote{Commonwealth Endeavour Scholar} and
P. D. Jarvis\footnote{Alexander von Humboldt Fellow}${}^,$\footnote{Tasmanian Senior Fulbright Scholar}

\vspace{0.3cm}

{\em School of Mathematics and Physics}\\
{\em University of Tasmania, Private Bag 37}\\
{\em Hobart, Tasmania 7001, Australia }\\
{\em E-mail: {\tt solj@utas.edu.au},  {\tt Peter.Jarvis@utas.edu.au}}

\thispagestyle{empty}
\begin{abstract}
\noindent
We study the relationship between one-dimensional fermion gas-impurity models and quantum dissipative systems, via the method of constructive bosonisation and unitary transformation. Starting from an ÔanisotropicÕ Coqblin-Schrieffer model, a new, exactly solvable, three-level quantum dissipative system is derived as a generalisation of the standard spin-$\textstyle{\frac 12}$ spin-boson model. The new system has two environmental oscillator baths with ohmic coupling, and admits arbitrary detuning between the three levels. All tunnelling matrix elements are equal, up to one complex phase which is itself a function of the longitudinal and transverse couplings in the integrable limit. Our work underlines the importance of re-examining the detailed structure of fermion-gas impurity models and spin chains, in the light of connections to models for quantum dissipative systems.\end{abstract}

\noindent

\vspace{10pt}

\end{centering} 

\vspace{125pt}

\end{titlepage}

\setcounter{footnote}{0}
\section{Introduction}
\label{sec:Introduction}
The understanding of how quantum systems behave in complex environments continues to be an important fundamental problem, with applications ranging from quantum computation and quantum information, to nano-devices and biological systems at the molecular level. Many insights into the system dynamics, thermodynamics, critical behaviour and entanglement properties can be gained by investigating simplified models of such quantum dissipative systems (QDS) \cite{Weiss2008, CostiandZarand1999, CostiandMcKenzie2003, Takahashi1999} with analytical techniques. A poignant example is provided by the well known equivalence between the spin-$\textstyle{\frac 12}$ anisotropic Kondo model (AKM) \cite{Kondo1964, CostiandZarand1999} and a particular case of the so-called spin-boson model \cite{Leggettetal1987} -- or two-level quantum dissipative system \cite{Leggettetal1987, CostiandZarand1999}. In this paper we propose a new model, providing an extension of this correspondence to a three-level QDS, also derivable via the method of constructive bosonisation \cite{Haldane1981, vonDelftandSchoeller1998} and mapping from an exactly-solvable, one-dimensional fermion gas-impurity system. The extension  is achieved by bosonisation of a fermion gas model with three-component fermions, starting with an \textit{anisotropic} form of a Coqblin-Schrieffer model \cite{CoqblinSchrieffer1969}. We demonstrate the exact solvability of this model, in the context of the standard analysis via coordinate wavefunctions, by verifying that the scattering data can be parametrised by an $R$-matrix of standard trigonometric type. As a three-level QDS, the model admits arbitrary detuning between the three energy levels, but the bath couplings take on special values in the integrable limit. Moreover, all tunnelling matrix elements are equal, up to one complex phase which is itself a function of the longitudinal and transverse couplings in the integrable limit. In section \ref{sec:ThreeLevel} we propose the new Hamiltonian, and outline the method of bosonisation and mapping in subsections \S\S \ref{subsec:Bosonisation} and \ref{subsec:UnitaryTransformation}, leading to a new three-level dissipative system model whose features we discuss. In \S \ref{subsec:Solvability} we show that the model is exactly solvable, and the final \S \ref{sec:Discussion} includes some comments on potential applications, and indicates directions for further work as concluding remarks. Remarks on notational details and some technical derivations are given in the appendix, \S A.

\section{Three-level quantum dissipative system }
\label{sec:ThreeLevel}
We wish to extend the well known \cite{Leggettetal1987, CostiandZarand1999} equivalence between the standard ($XXZ$-type), spin-$\textstyle{\frac 12}$ AKM model, and a particular case of the 
spin-$\textstyle{\frac 12}$ spin-boson model - the simplest possible two-level QDS \cite{Leggettetal1987}. In order to motivate and present the extension, we firstly briefly introduce these two models and their parameters.
In second-quantised form, the AKM Hamiltonian is
\begin{align}
\label{eq:HAKM}
H_{AKM} = & \, \sum_{p} \hbar v_Fp \big(\mbox{\boldmath{$:$}}c^\dagger_{p \uparrow}c_{p \uparrow}+c^\dagger_{p \downarrow}c_{p \downarrow}\mbox{\boldmath{$:$}}\big) +
J_\parallel \displaystyle{\sum_{p,p'}} \big(\mbox{\boldmath{$:$}}c^\dagger_{p \uparrow}c_{p' \uparrow} \! - \!
c^\dagger_{p \downarrow}c_{p' \downarrow}\mbox{\boldmath{$:$}}\big)S_z  
 + \frac12 J_\perp\displaystyle{\sum_{p,p'}} \big(c^\dagger_{p \downarrow}c_{p' \uparrow}
S_+ \! + \!
c^\dagger_{p \uparrow}c_{p' \downarrow}S_-\big) \nonumber \\
\equiv & \,
\sum_{p, \alpha}  \hbar v_Fp \mbox{\boldmath{$:$}}c^\dagger_{p \alpha}c_{p \alpha}\mbox{\boldmath{$:$}}\,
+ \, J_\parallel \!\!  \!\! \displaystyle{\sum_{p,\alpha,p',\alpha'}} \mbox{\boldmath{$:$}} c^\dagger_{p \alpha}(\sigma_z)_{\alpha \alpha'}c_{p' \alpha'}\mbox{\boldmath{$:$}}S_z\, +
\nonumber \\ 
& \, \qquad \qquad \qquad \qquad \qquad \qquad
 + \frac12 J_\perp  \!\! \!\!\displaystyle{\sum_{p,\alpha,p',\alpha'}}\big( c^\dagger_{p \alpha}(\sigma_-)_{\alpha \alpha'} c_{p' \alpha'}S_+ +
c^\dagger_{p \alpha}(\sigma_+)_{\alpha \alpha'} c_{p' \alpha'}S_- \big).
\end{align}
Here, $c^\dagger_{p \alpha}$, $c_{p \alpha}$ are respectively creation and annihilation operators for spin-$\textstyle{\frac 12}$ electrons with wavenumber $p$, and energy $\hbar v_F p$ linearised about the Fermi level ($p = 2\pi n_p/L$, $n_p \in {\mathbb Z}$, for system length $L$ and periodic boundary conditions), while $\alpha, \beta = 1,2$ or $\uparrow, \downarrow$ are the spin orientations. The second form of (\ref{eq:HAKM}) is given to facilitate comparisons with the three-level extension of the model, where Gell-Mann matrices will take the role of the Pauli matrices. The symbol $\mbox{\boldmath{$:$}} \cdots \mbox{\boldmath{$:$}}$ stands for operator normal ordering relative to the Fermi sea, the vacuum state $|0\rangle$ annihilated by both $c_p$ ($p>0$) and $c^\dagger_p$ ($p\le0$). The electrons interact with a localised magnetic impurity atom (taken to be at the origin in coordinate space) via (anisotropic) coupling between the spin density at the origin with the impurity spin. For clarity its spin operators are written as $S_z$, $S_\pm \equiv S_x \pm i S_y$ rather than as Pauli matrices $\textstyle{\frac 12}\mbox{\boldmath{$\sigma$}}$. Finally, to (\ref{eq:HAKM}) can be added an external magnetic interaction of the form $h S_z$, $h := \mu \mu_B B $ for impurity magnetic
moment $\mu$.

The spin-$\textstyle{\frac 12}$ spin-boson, or two-level QDS, Hamiltonian is \cite{Leggettetal1987}: 
\begin{align}
\label{eq:HSB}
H_{SB} = & \, 
\textstyle{\frac 12} \varepsilon \sigma_z - \textstyle{\frac 12} \hbar\Delta \sigma_x + \sum_i \hbar \omega_i b^\dagger_i b_i  + 
\sigma_z \sum_i \hbar C_i(b_i + b_i^\dagger).
\end{align}
In (\ref{eq:HSB}), the two-level system has detuning (level- or well-asymmetry) parameter $\varepsilon$, and tunnelling amplitude $\Delta$ between the two levels or wells.
The index $\mbox{}_i$ labels a set of harmonic oscillators playing the role of the environment or `bath', with energies determined by the associated frequencies $\omega_i $ (the zero-point contributions $\textstyle{\frac 12} \hbar \omega_i$ have been removed by an appropriate shift). The bath interactions with the two-level system, with strengths given by the coupling constants $C_i$, affect the energy in the upper and lower levels (which are eigenstates of $\sigma_z$). The overall influence of the oscillators on the reduced system is described by the Ôspectral densityÕ $J(\omega): = \sum_i C_i^2 \delta(\omega - \omega_i)$ \cite{CaldeiraandLeggett1983}. For the cases of interest, it has the so-called \emph{ohmic} form, $J(\omega) \, \propto \, \omega\hspace{.3ex} e^{-(\omega/\omega_c)}$, where $\omega_c$ is the frequency cutoff. An important implicit parameter which is critical to the behaviour of the QDS is the ohmic coupling $\alpha$, the proportionality constant which determines the strength of this relationship between $J(\omega)$ and $\omega$
(which is of course linear, for $\omega \ll \omega_c$).

It is well known that the AKM model (\ref{eq:HAKM}), which is also closely related to the $XXZ$ Heisenberg spin chain, is exactly solvable \cite{Baxter1982,TsvelickandWiegmann1983}, and so this in-principle complete analytical access to all details of the spectrum, eigenstates and correlation functions is conferred, through the transcription via bosonisation and an associated unitary transformation, on the spin-boson model itself. In this paper we expand on the dictionary of such exactly-solvable fermion gas systems, which may be brought into QDS form, presenting a new, \emph{three-level} QDS. Starting with an appropriate 1$D$ fermion gas model, we reiterate the steps of the standard recipe for the bosonisation mapping and unitary transformation which established the equivalence between (\ref{eq:HAKM}) and (\ref{eq:HSB}). Finally, we verify using the standard coordinate space approach that the starting model is indeed in the exactly-solvable class, thus underlining the utility of the new three level QDS model. 

The starting point is an equivalent of the spin-$\textstyle{\frac 12}$ AKM, but for \emph{three} component fermions rather than spin-$\textstyle{\frac 12}$. In magnetic systems, multicomponent fermions find applications in the Coqblin-Schrieffer (C-S) model \cite{CoqblinSchrieffer1969}; for the transcription to a three-level QDS, we shall require in the fermionic picture, an extended parametrisation of the interaction between the local and impurity `spins' with additional terms of `anisotropic'  type: an `AC-S' model. The full Hamiltonian thus contains the free-fermion kinetic term, external ÔmagneticÕ field interactions, and the analogue of both transverse and longitudinal interaction terms between the localised fermion `spin' operators at the origin, and those of the impurity atom:  
\begin{align}
\label{eq:HACS}
H^F_{ACS} = & \,
\sum_{p, \alpha=1}^3  \hbar v_Fp \,\mbox{\boldmath{$:$}}c^\dagger_{p \alpha}c_{p \alpha}\mbox{\boldmath{$:$}} + \sum_\alpha h_\alpha S_{\alpha \alpha}
+\displaystyle{\sum_{p,p',\alpha}} J_\parallel \,\mbox{\boldmath{$:$}} c^\dagger_{p \alpha}c_{p' \alpha}\mbox{\boldmath{$:$}}S_{\alpha \alpha} +
J_\perp  \!\! \!\! \displaystyle{\sum_{p,p', \alpha<\beta}}\big( e^{i\zeta_{\alpha\beta}}
c^\dagger_{p \beta} c_{p' \alpha}S_{\alpha \beta} \!+\! \mbox{h.c.} \big).
\end{align}
Here $p$ is the fermion wavenumber as before; $\alpha, \beta = 1,2,3$ label the \emph{three} independent components. The magnetic impurity operators $S_{\alpha \beta}$ are generators of the $SU(3)$ Lie algebra for $\alpha,\beta = 1,2,3$, provided $S_{11}+S_{22}+S_{33} = 0$; more generally it will be convenient to drop this condition and regard the 9 \emph{independent} operators $S_{\alpha \beta}$ as generators of $U(3)$.
In the course of the bosonisation transcription of the model, it will turn out that the complex phases $\zeta_{\alpha \beta}$, included for generality in the first instance, are all equal, and in fact parametrised in terms of the $J_\parallel$ and $J_\perp$ couplings. 
 As we shall see, these points will be of significance in the final QDS version, (\ref{eq:HBQDS}). We now turn to a brief discussion of the technicalities of this reformulation.

\subsection{Bosonisation}
\label{subsec:Bosonisation}
The key element in the famous fermion-boson correspondence \cite{Leggettetal1987, vonDelftandSchoeller1998} is the recognition that, for an infinite number of fermionic modes of species $\alpha$, the following bilinear combinations,
\begin{align}
b^\dagger_{k \alpha}  = & \, i\sqrt{  {\frac{2\pi}{Lk}}}\sum_{p=-\infty}^{\infty}\mbox{\boldmath{$:$}} c^\dagger_{p\!+\!k\, \alpha} c_{p\alpha}\mbox{\boldmath{$:$}}, \qquad 
b_{k \alpha}  = - i\sqrt{  {\frac{2\pi}{Lk}}}\sum_{p=-\infty}^{\infty} \mbox{\boldmath{$:$}} c^\dagger_{p\!-\!k \, \alpha} c_{p\alpha} \mbox{\boldmath{$:$}}
 \nonumber
 \end{align}
fulfil the Heisenberg commutation relations for an infinite set of bosonic modes, namely
${[}b_{k\alpha}, b^\dagger_{k'\beta} {]} = \delta_{k k'}\delta_{\alpha\beta}$, and ${[}b_{k\alpha},b_{k' \beta}{]} = 0 = {[} b^\dagger_{k\alpha},b^\dagger_{k' \beta} {]}$, for $k = 2\pi n_k/L$, $n_k = 1,2,3,\cdots$. Building the appropriate multicomponent local quantum fields in one dimension, 
\begin{align}
\label{eq:PsiPhiNoA}
\psi_\alpha(x) = & \, \sqrt{  {\frac{2\pi}{L}}} \sum_p e^{-ipx}c_{p\alpha},
\qquad \psi^\dagger_\alpha(x) =  \sqrt{  {\frac{2\pi}{L}}} \sum_p e^{ipx}c^\dagger_{p\alpha},
\nonumber \\
\varphi_{\alpha}(x) = & \, -\sum_{k>0} \sqrt{  {\frac{2\pi}{Lk}}} 
                  \big(e^{-ikx}  b_{k \alpha} \!+\! e^{ikx} b^\dagger_{k \alpha} \big)
\end{align}
yields the identification at the level of operators on Fock space
\begin{align}
\label{eq:OperatorIdNoA}
\psi_\alpha(x) = & \, \sqrt{  {\frac{2\pi}{L}}}  {\mathcal F}_\alpha \,\mbox{\boldmath{$:$}} 
e^{-i\varphi_\alpha(x)}\mbox{\boldmath{$:$}}
\end{align}
where the prefactors ${\mathcal F}_\alpha$, the so-called Klein operators, must fulfil certain additional relations to retain the anti-commutation relations necessary for fermionic operators (see below).

It is evident from (\ref{eq:HAKM}) and (\ref{eq:HACS}) above, that the necessity to work with an infinite number of fermionic modes implies that the single particle dispersion relation is extrapolated indefinitely above and below the fermi level. As a consequence, the energy spectrum of the model as a whole is formally unbounded below. In practice this situation is dealt with by introducing a momentum cutoff, which is adequate for most situations in condensed matter. However, in the present context it is crucial to maintain the rigorous mathematical fermion-boson correspondence and isomorphism of Hilbert spaces throughout, and so the regularisation of `constructive bosonisation' is adopted \cite{vonDelftandSchoeller1998}. This introduces a regularisation parameter $a\rightarrow 0$ which sets a scale for the suppression of contributions from wavenumbers $|p| \gtrsim a^{-1}$ away from the fermi surface, by modifying (\ref{eq:PsiPhiNoA}) above to 
\begin{align}
\varphi_{\alpha}(x) = & \, -\sum_{k>0} \sqrt{  {\frac{2\pi}{Lk}}} 
                  \big(e^{-ikx}  b_{k \alpha} \!+\! e^{ikx} b^\dagger_{k \alpha} \big){{ e^{-ak/2} }}.
\nonumber
\end{align}                 
An important consequence is that normal ordering in operator products can be re-expressed in terms of ordinary products in an expansion in powers of $a$. In particular, (\ref{eq:OperatorIdNoA}) becomes
\begin{align}
\label{eq:OperatorIdWithA}
\psi_\alpha(x) = & \, \lim_{a\rightarrow 0}\left( \frac{{\mathcal F}_\alpha}{\sqrt{a}}  e^{-i\varphi_\alpha(x)}\right).
\end{align}
With these definitions in hand, we can proceed to develop the bosonic counterparts of the various terms in (\ref{eq:HACS}) in order to expose the structure of the three level QDS equivalent.

It should be noted that (\ref{eq:OperatorIdNoA}), (\ref{eq:OperatorIdWithA}) are only valid as operator identities when acting on the zero fermion number sectors of the respective fermionic Hilbert spaces. The corrected expressions should have additional charge-dependent phase factors ${\exp(2 \pi i x/L)}^{N_\alpha}$ for charge $N_\alpha$ in each case, which in turn can be seen as deriving from the equivalent formula entailing the number operator $\widehat{N}_\alpha$,
\begin{align}
\widehat{N}_\alpha = & \, \sum_p \mbox{\boldmath{$:$}} c^\dagger_{p\alpha} c_{p\alpha} \mbox{\boldmath{$:$}},
\nonumber
\end{align}
applied to each charge eigenspace.
Most of the steps in the bosonisation transcription entail expressions which are bilinear in fermions, and for field quantities evaluated locally at the magnetic impurity ($x=0$), so these phases tend to cancel. However, $\widehat{N}_\alpha$-dependent terms do occur, and their treatment will be taken up in the discussion of the final QDS model below, and technical remarks relegated to the appendix, \S A.2.

\setcounter{equation}{0}
\renewcommand{\theequation}{$H^B_{(\roman{equation})}$ }
The systematics by which the couplings and modes are reorganised can be seen by inspecting the `magnetic' term:
\begin{align}
\label{eq:HBmagnetic}
& \, \sum_{\alpha=1}^3 h_\alpha S_{\alpha \alpha}
\equiv h_{\texttt 0} S_{\texttt 0} + h_{\texttt 3} S_{\texttt 3} +h_{\texttt 8} S_{\texttt 8},
\end{align}
entailing a relabeling from diagonally or doubly-indexed quantites ${x}_{\alpha \beta}, \alpha, \beta=1,2,3$ to the new set ${x}_{\texttt A}, {\texttt A} = {\texttt 3},{\texttt 8},{\texttt 0}$, (reserving  ${\texttt A}=
 {\texttt 1},{\texttt 2},{\texttt 4},{\texttt 5},{\texttt 6},{\texttt 7}$ for off-diagonal labels), and using  standard Jacobi three-body combinations:
\[
x_{\texttt 3} = \textstyle{\frac{1}{\sqrt{2}}   }(x_{11}-x_{22}),
\quad x_{\texttt 8} = \textstyle{\frac{1}{ \sqrt{6}}  }(x_{11}+x_{22}-2 x_{33}),
\quad  x_{\texttt 0} = \textstyle{\frac{1}{\sqrt{3}}  }(x_{11}+x_{22}+ x_{33}).
\]
The kinetic term is similarly expanded (up to fermion number-dependent terms) as
\[ 
\displaystyle{\sum_{p, \alpha=1}^3} \hbar v_F p\,\mbox{\boldmath{$:$}} c^\dagger_{p \alpha} c_{p \alpha}\mbox{\boldmath{$:$}}  = 
\displaystyle{\sum_{k>0}}  \hbar v_F k\, \big( b^\dagger_{{\texttt 3}k}b_{{\texttt 3}k} + b^\dagger_{{\texttt 8}k}b_{{\texttt 8}k} +b^\dagger_{{\texttt 0}k}b_{{\texttt 0}k} \big).
\]
The transverse `spin' interaction terms remain off-diagonal, and are not affected by normal ordering, leading to\footnote{Note that a common transcription of the standard two-level Kondo/spin-boson equivalence uses Wannier operator notation, where the coupling constant dimensions are scaled by a factor proportional to the system size (see for example \cite{Leggettetal1987}).}
\[
\frac{L}{2\pi a}J_\perp \displaystyle{\sum_{\alpha<\beta}} \left( e^{i\zeta_{\alpha\beta}}e^{-i(\varphi_\alpha(0) 
-\varphi_\beta(0))}{\mathcal F}^\dagger_{\beta}{\mathcal F}_{\alpha} S_{\alpha \beta} +  e^{-i\zeta_{\alpha\beta}}e^{i(\varphi_\alpha(0) -\varphi_\beta(0))}{\mathcal F}^\dagger_{\alpha}{\mathcal F}_{\beta} S_{\beta \alpha} \right).
\]
By contrast, the longitudinal `spin' couplings (with diagonal fermion bilinears) simply become combinations of the oscillator modes themselves when the bosonisation is implemented, in the form
\begin{align}
\label{eq:HBlongitudinal}
J_\|  \! \displaystyle{\sum_{p,p',\alpha}} \!
 \mbox{\boldmath{$:$}}c^\dagger_{p \alpha} c_{p' \alpha}\mbox{\boldmath{$:$}}S_{\alpha \alpha}
= & \, 
J_\parallel  \displaystyle{\sum_{k>0}} \sqrt{\frac{kL}{2\pi}}e^{-ka/2} \,
i\!\left(S_{\texttt 3}(b_{{\texttt 3}k}\!-\!b^\dagger_{{\texttt 3}k})+S_{\texttt 8}(b_{{\texttt 8}k}\!-\!b^\dagger_{{\texttt 8}k})+S_{\texttt 0}(b_{{\texttt 0}k}\!-\!b^\dagger_{{\texttt 0}k})\right)
\end{align}
-- again together with additional terms proportional to fermion-number.

\subsection{Unitary transformation}
\label{subsec:UnitaryTransformation}
These contributions are aggregated together with an additional transformation, a conjugation $U \cdot U^{-1}$ by the operator
\[
U = \exp(\displaystyle{i {\textstyle{\sum_\alpha} \varphi_\alpha(0) S_{\alpha \alpha} }}) \equiv
\exp(\displaystyle{i(\varphi_{\texttt 3}(0) S_{\texttt 3} +\varphi_{\texttt 8}(0) S_{\texttt 8} + \varphi_{\texttt 0}(0) S_{\texttt 0} )}).
\]
It is evident from the commutation relations of the $U(3)$ Lie algebra, ${[}S_{\alpha\alpha}, S_{\alpha \beta}{]} = S_{\alpha \beta}$, ${[}S_{\beta\beta}, S_{\alpha \beta}{]} = - S_{\alpha \beta}$ (with $\alpha \ne \beta$) that this unitary transformation will \emph{cancel} the offending scalar exponentials in the transverse coupling terms, leaving the composite operators
\begin{align}
\label{eq:HBtransverse}
\frac{L}{2\pi a}J_\perp \displaystyle{\sum_{\alpha<\beta}} \left( e^{i\zeta_{\alpha\beta}}{\mathcal F}^\dagger_{\beta}{\mathcal F}_{\alpha} S_{\alpha \beta} +e^{-i\zeta_{\alpha\beta}} {\mathcal F}^\dagger_{\alpha}{\mathcal F}_{\beta} S_{\beta \alpha} \right);
\end{align}
the kinetic terms acquire an additional commutator contribution of the same structure as the longitudinal coupling terms, which themselves commute with $U$:
\begin{align}
\label{eq:HBkinetic}
U \, \displaystyle{\sum_{k>0}}  &\hbar v_F k  \big( b^\dagger_{{\texttt 3}k}b_{{\texttt 3}k} \!+\! b^\dagger_{{\texttt 8}k}b_{{\texttt 8}k}  \!+\!b^\dagger_{{\texttt 0}k}b_{{\texttt 0}k} \big) \, U^{-1}  
= \nonumber \\[-.1cm]
& \,\displaystyle{\sum_{k>0}}  \hbar v_F k  \big( b^\dagger_{{\texttt 3}k} b_{{\texttt 3}k}  \!+\!
 b^\dagger_{{\texttt 8}k}b_{{\texttt 8}k}  \!+\!b^\dagger_{{\texttt 0}k}b_{{\texttt 0}k} \big) 
- \hbar v_F  \displaystyle{\sum_{k>0}} \sqrt{\frac{2\pi k}{L}}e^{-ka/2}\,
i\! \left(S_{\texttt 3}(b_{{\texttt 3}k}\!-\!b^\dagger_{{\texttt 3}k})+S_{\texttt 8}(b_{{\texttt 8}k}\!-\!b^\dagger_{{\texttt 8}k})+S_{\texttt 0}(b_{{\texttt 0}k}\!-\!b^\dagger_{{\texttt 0}k})\right) 
\end{align}
(a further term arising from the double commutator in the conjugation by the exponential yields a power series in $a$ whose sum can be removed as an additional overall constant).
The outcome of the transcription of the AC-S Hamiltonian (\ref{eq:HACS}) is thus the combination of the above 
`magnetic', longitudinal, transverse and kinetic terms:
\[
U \cdot H_{ACS} \cdot U^{-1} = H^B_{(i)}+H^B_{(ii)}+H^B_{(iii)}+H^B_{(iv)}.
\]

\setcounter{equation}{6}
\renewcommand{\theequation}{\arabic{equation}}
The reinterpretation of the right-hand side as a dissipative system Hamiltonian proceeds by consideration of the composite operators in \ref{eq:HBtransverse} which involve the off-diagonal generators $S_{\alpha \beta}$, $\alpha \ne \beta$ of $U(3)$ in combination with Klein operators. It is easily checked using the algebraic properties \cite{vonDelftandSchoeller1998}
\[
 {\mathcal F}_\alpha {\mathcal F}^\dagger_\alpha =  {\mathcal F}^\dagger_\alpha {\mathcal F}_\alpha =1, 
 \qquad 
 {\{} {\mathcal F}^\dagger_\alpha,{\mathcal F}_\beta{\}} = {\{} {\mathcal F}_\alpha, {\mathcal F}_\beta{\}} =
 {\{} {\mathcal F}^\dagger_\alpha,{\mathcal F}^\dagger_\beta{\}} = 0, \quad \alpha \ne \beta
 \]
that, provided the $S_{\alpha \beta}$ are indeed \emph{elementary} $3\! \times \!3$ matrices, the composite
operators defined by $S'{}_{\alpha \beta} := -{\mathcal F}{}^\dagger_\beta {\mathcal F}_\alpha S{}_{\alpha \beta}$, $S'{}_{\alpha \alpha} := {\mathcal F}{}^\dagger_\alpha {\mathcal F}_\alpha S{}_{\alpha \alpha} = S_{\alpha \alpha}$  fulfil the usual $U(3)$ commutation relations, and can be identified with operators acting between the states of the quantum three-level system in the QDS interpretation.

The next step is to combine $H^B_{(ii)}$ and $H^B_{(iv)}$, with the recognition that $S_{\texttt 0}$ is the linear Casimir invariant of $U(3)$ (and is certainly proportional to the $3\! \times \!3$ identity matrix if the original operators are represented with elementary matrices). Thus the terms involving $b_{{\texttt 0}k}$ and $b^\dagger{}_{{\texttt 0}k}$ are entirely quadratic and linear -- completing the square for such `displaced oscillator' modes enables their contributions to be combined, up to an (infinite) shift in the energy, into a sum of kinetic energy terms for an infinite set of oscillator modes which do not interact with the remainder of the system and can be dropped from the final model. By the same token, the $h_{\texttt 0} S_{\texttt 0}$ term can be dropped from $H^B_{(i)}$.

In order to emphasize the similarity between the two-level QDS, the spin-$\textstyle{\frac 12}$ spin-boson model (\ref{eq:HSB}), and the new system, we adopt the standard $3\! \times \!3$ Gell-Mann matrices $\lambda_{\texttt A}$, ${\texttt A} = {\texttt 1},\cdots, {\texttt 8}$ as an orthogonal basis for the $SU(3)$ generators in the fundamental representation, to play the role of the Pauli matrices in (\ref{eq:HAKM}). Gathering all terms, the form of the three level QDS Hamiltonian $U \cdot H_{ACS} \cdot U^{-1}  \rightarrow H^B_{QDS}$ becomes finally 
\begin{align}
\label{eq:HBQDS}
H^B_{QDS} := & \, \varepsilon_{\texttt 3} \lambda_{\texttt 3}+\varepsilon_{\texttt 8} \lambda_{\texttt 8}+
     \Delta (\lambda_{\texttt 1} \!+\!\lambda_{\texttt 4}  \!+\! \cos\zeta \lambda_{\texttt 6}+ \sin\zeta  \lambda_{\texttt 7}) + 
     {\sum}_{k} \hbar \omega_k (b^\dagger_{k{\texttt 3}} b_{k{\texttt 3}} \!+\! b^\dagger_{k{\texttt 8}} b_{k{\texttt 8}})  + \nonumber \\
     & \, + {\sum}_k \hbar C_{{\texttt 3}k} \lambda_{\texttt 3}(b_{k{\texttt {\texttt 3}}} + b^\dagger_{k{\texttt {\texttt 3}}})+
     \hbar C_{{\texttt 8}k} \lambda_{\texttt 8}(b_{k{\texttt {\texttt 8}}} + b^\dagger_{k{\texttt {\texttt 8}}}). 
\end{align}

The QDS parameters have the following definitions in terms of those of the original AC-S model. From above, the detuning parameters $\varepsilon_{\texttt 3}$ and $\varepsilon_{\texttt 8}$ are simply $h_{\texttt 3}$ and $h_{\texttt 8}$ respectively, and from the kinetic terms the oscillator baths have frequency spectrum $\omega_k = v_F k$ provided
$\omega \ll \omega_c$, where $\omega_c$ is the cutoff frequency $\omega_c = v_F/a$. 
The tunnelling matrix elements are given in terms of the transverse coupling strength of the original model, $\Delta \equiv -J_\perp L /2\pi a$, modulated by a complex phase. By an appropriate basis choice, this phase may be shifted on to the $2,3$ sector, and expressed in the orthogonal basis by a combination of the corresponding Gell-Mann matrices, namely $\lambda_{\texttt 6}$ and $\lambda_{\texttt 7}$, rotated by angle $\zeta := \zeta_{23} - \zeta_{13} + \zeta_{12}$ (see (\ref{eq:HACS}) and also (\ref{eq:Scalc}), (\ref{eq:TransfR}) below).

The dissipative terms have been re-written in the conventional coordinate-coupled form 
$\cong (b\!+\!b^\dagger)$ rather than the imaginary (momentum) combinations $i(b\!-\!b^\dagger)$ appearing in the above derivation  by means of a canonical transformation $b^\dagger \rightarrow -i b^\dagger$, $b \rightarrow  i b$. Evidently, the overall dissipative couplings are a combination of contributions from different terms, although both coefficients $C_{{\texttt 3}k}$ and $C_{{\texttt 8}k}$ are equal:
\begin{align}
\label{eq:Cdefn}
C_{{\texttt 3}k}=&\,C_{{\texttt 8}k}\equiv C_k = - v_F \sqrt{\frac{2\pi k}{L}}e^{-\omega_k/2\omega_c}
\left( 1 - \frac{J_\parallel L}{2\pi \hbar v_F}\right).
\end{align}
The spectral frequency $J(\omega)$ follows directly from the definition (in the limit $a\rightarrow 0$). As shown explicitly in \S \ref{subsec:OhmicCoupling}, this has the ohmic form
\begin{align}
\label{eq:JnAlphaDefn}
J(\omega) =&\, \alpha\, \omega e^{\displaystyle{-\omega/\omega_c}}, \qquad
\mbox{where}\qquad \alpha := \left( 1 - \frac{J_\parallel L}{2\pi \hbar v_F}\right)^{\!\!2}.
\end{align}
The additional parameters emerging from the details of the way the AC-S model and its bosonisation are implemented are thus the cutoff frequency $\omega_c$ and the dimensionless ohmic coupling $\alpha$ (not to be confused with the spin-label $\alpha$).

As mentioned above, the fermion-number dependence of the bosonisation transcription still requires explanation. Indeed, the introduction of the Klein-factor dependent operators $S'_{\alpha\beta}$ as effective $U(3)$ generators implies that the three states of the quantum system in fact lie across different charge sectors. This situation, and at the same time the treatment of the residual fermion-number dependent terms, is resolved by noting that the original model (\ref{eq:HACS}) has three conserved quantum numbers
$\widehat{N}_\alpha + S_{\alpha \alpha}$, $\alpha = 1,2,3$ or 
$\widehat{N}_{\texttt 3} + S_{{\texttt 3}}$, $\widehat{N}_{\texttt 8} + S_{{\texttt 8}}$, and $\widehat{N}_{\texttt 0} + S_{{\texttt 0}}$ in terms of relative degrees of freedom. Of course $S_{\texttt 0}$ is proportional to the identity matrix, and so a projection onto an eigenspace with fixed eigenvalue $M_{\texttt 0}$ is tantamount to 
fixing the total fermion number at say $N_{{\texttt 0}}$ which is certainly a conserved quantity. The 
system further admits a projection onto fixed eigenspaces of the remaining two operators with eigenvalues $M_{\texttt 3}$ and $M_{\texttt 8}$, say. As shown in \S \ref{subsec:Projection}, these projections leave the form of (\ref{eq:HBQDS}) unchanged. However, the detuning parameters $\varepsilon_{\texttt 3}, \varepsilon_{\texttt 8}$ need to be shifted from their orignial values $h_{\texttt 3}, h_{\texttt 8}$ to absorb additional $M_{\texttt 3}$- and $M_{\texttt 8}$-dependent contributions.

\subsection{Exact solvability and extensions of the model }
\label{subsec:Solvability}
The equivalence of the models (\ref{eq:HACS}) and (\ref{eq:HBQDS}) establishes that the three component fermi gas model does indeed have a dissipative system counterpart. The utility of this observation of course derives from also showing that the starting model belongs to the exactly solvable class. The standard coordinate analysis, or an equivalent algebraic formulation in the context of the associated spin chain, requires that the model admit an $R$-matrix with the appropriate properties. In the present case, following \cite{TsvelickandWiegmann1983}, we require that the single particle-impurity scattering matrix ${\sf S}$, expressible as the exponential of the interaction component of the Hamiltonian $H_{int}(J_\parallel, J_\perp)$,
can be reparametrised in terms of the $R$-matrix $R(x^{\alpha=1},q)$ for some arbitrary but fixed value of the (additive) spectral parameter, say $\alpha =1$.  Thus we demand
\[
{\sf S} = e^{iH_{int}(J_\parallel, J_\perp)} \equiv R(x^{\alpha=1},q)
\]
in such a way that the parameters $x,q$ become functions of the couplings $J_\parallel, J_\perp$.
We proceed by an explicit evaluation of ${\sf S}$. From (\ref{eq:HACS}) we have, using elementary $3\! \times \!3$ matrices $e_{\alpha \beta}$,
\begin{align}
\label{eq:Scalc}
H_{int} = & \, J_\parallel \sum_{\alpha} e_{\alpha\alpha}\otimes e_{\alpha\alpha} +
J_\perp \sum_{\alpha<\beta} \big( e^{i\zeta_{\alpha\beta} } e_{\alpha\beta}\otimes e_{\beta\alpha}\!+\! e^{-i\zeta_{\alpha\beta} } e_{\beta\alpha}\otimes e_{\alpha\beta}\big), \nonumber \\
\mbox{so}\qquad \quad
 {\sf S} = & \, e^{iJ_\parallel}\sum_{\alpha} e_{\alpha\alpha}\otimes e_{\alpha\alpha}+ \cos J_\perp\sum_{\alpha\ne \beta} e_{\alpha\alpha}\otimes e_{\beta\beta} + i \sin J_\perp  \sum_{\alpha<\beta} \big( e^{i\zeta_{\alpha\beta} } e_{\alpha\beta}\otimes e_{\beta\alpha}\!+\! e^{-i\zeta_{\alpha\beta} } e_{\beta\alpha}\otimes e_{\alpha\beta}\big).
\end{align}
This must be compared with the known forms \cite{Jimbo1986,Arnaudonetal2006} (see also \cite{Bazhanov1985}) for standard trigonometric $R$-matrices of the appropriate dimension,
\begin{align}
\label{eq:ArnaudonJimbo}
R(x,q) = & \, (qx - q^{-1}x^{-1}) \sum_{\alpha} e_{\alpha\alpha}\otimes e_{\alpha\alpha}   + 
(x - x^{-1}) \sum_{\alpha\ne \beta} e_{\alpha\alpha}\otimes e_{\beta\beta} + \nonumber 
\\
& \, + (q - q^{-1})\sum_{\alpha<\beta} \big(x e_{\alpha\beta}\otimes e_{\beta\alpha} + x^{-1} e_{\beta\alpha}\otimes e_{\alpha\beta}\big).
\end{align}
This expression clearly has the correct structure to be identified with the scattering matrix ${\sf S}$ if 
$\zeta_{12}$ $= \zeta_{13}$ $=\zeta_{23}$ with phase factors identified with $x$. Adopting logarithmic parameters $x=e^{i\overline{f}}$, $q=e^{\overline{\mu}}$, thus with $\zeta \equiv \overline{f}$, the $R$-matrix is up to a factor of 2,
\begin{align}
\label{eq:TransfR}
R(x,q) = & \, \sinh(i\overline{f}+ \overline{ \mu}) \sum_{\alpha} e_{\alpha\alpha}\otimes e_{\alpha\alpha}   + 
i\sin(\overline{f}) \sum_{\alpha\ne \beta} e_{\alpha\alpha}\otimes e_{\beta\beta} + \nonumber 
\\
& \, + \sinh(\overline{\mu})\sum_{\alpha<\beta} \big(e^{i\overline{f}} e_{\alpha\beta}\otimes e_{\beta\alpha} +e^{-i\overline{f}} e_{\beta\alpha}\otimes e_{\alpha\beta}\big). 
\end{align}
Comparing the ratios of coefficients in the expressions (\ref{eq:TransfR}), (\ref{eq:Scalc}) leads directly to the reparametrisation of $\overline{f}$, $\overline{\mu}$ in terms of $J_\parallel$, $J_\perp$:
\begin{align}
\label{eq:Parametrisations}
\cosh \overline{\mu} = & \, \frac{\cos J_\parallel}{\cos J_\perp}; \qquad
\cot^2 \overline{f} =  \frac{\sin^2 J_\parallel}{\sin (J_\perp \!+\! J_\parallel)
                                          \sin ( J_\perp \!-\! J_\parallel) } .
\end{align}
With the three-level model admitting a reparametrisation showing equivalence to the exactly solvable trigonometric R-matrix, it is clear that the proposed model belongs to this rare and important class of exactly solvable dissipative systems.

\section{Discussion}
\label{sec:Discussion} 
In conclusion, this report has followed the constructive bosonisation approach to propose a new exactly solvable three-level quantum dissipative system. Although the background formalism is well known, we have presented concrete details and careful explanations in order to expose the technicalities of the required manipulations, and we anticipate that the methods of this paper may be deployed to find other exactly solvable quantum dissipative system models. In the present case it can be expected that further study will yield insights into the physics of this system as an instance of a QDS model, to be compared and contrasted with the already well studied mapping of the AKM to  the spin-$\textstyle{\frac 12}$ spin-boson model.

As a generalisation of the two-level spin-boson/Kondo model correspondence, the three-level analogue presented in this work belongs to the same family of related problems and models. In particular it is interesting to note the relationship between the present model and the triangular lattices and quantum Brownian motion discussed in \cite{Afflecketal2001}. It appears that the transverse field terms in (\ref{eq:HBQDS}) are equivalent to hops on this triangular lattice. The model in \cite{Afflecketal2001} is shown to correspond to the two-dimensional $3$-state Potts model with a boundary, with critical behaviour derivable through $c=2$ boundary conformal field theory, and it would be instructive to formalise the correspondence to the present three-level dissipative system. Furthermore, it has been pointed out that the present model bears connections to quantum wire junctions, the dissipative Hofstadter model and open string theory as presented in \cite{Chamonetal2003}. The current model contains further generalisations to these systems by including marginal operators coupled to the diagonal elements of the $SU(3)$ algebra.
 
Further examples of three-level system-environment models to which our new exactly solvable three-level QDS might be applied include three-level quantum dots, single qubit systems addressed by an extra ancillary state, or qutrit states, triatomic triple well potentials, such as ammonia ($NH_3$) and methyl ($-CH_3$), as well as Bose-Einstein condensate atomic transistors \cite{Stickneyetal2007}, which have a three well structure. To further develop the model one should resolve the full spectrum and eigenstates of the Hamiltonian via the Bethe \emph{Ansatz} \cite{Bethe1931}, allowing for calculation of dynamical and thermodynamical quantities of interest. One may also be interested in investigating the vacuum sector dependence \cite{vonDelftandSchoeller1998} and finite size effects \cite{ZarandandvonDelft2000} in the bosonisation. The study of entanglement between quantum systems and dissipative environments \cite{Amicoetal2008,OhandKim2006} may also be examined within this impurity-bath system. Generically, it is clear that our analysis of the details of the constructive bosonisation and unitary mapping technique suggests that, in the light of potential new applications to quantum dissipative systems, the well-known connections between fermion gas-impurity models and spin chains warrant re-examination.

\subsection*{Acknowledgements}
We thank Ross McKenzie for discussions in the early stages of this work and Gergely Zar\'and and the Referees for instructive suggestions and improvements to the contextualization of this work. This project was in part funded by the Commonwealth of Australia Endeavour Awards. 


\bibliographystyle{unsrt}

\begin{thebibliography}{widest-label}

\bibitem{Weiss2008}
U. Weiss.
\newblock Quantum Dissipative Systems, Third Edition.
\newblock {\em World Scientific Publishing}, 2008.

\bibitem{CostiandZarand1999}
T.A. Costi and G. Zar\'and.
\newblock Thermodynamics of the dissipative two-state system: {A} {B}ethe {A}nsatz study.
\newblock {\em Physical Review B}, 59(19):12398--12418, 1999.

\bibitem{CostiandMcKenzie2003}
T.A. Costi and R.H. McKenzie.
\newblock Entanglement between a qubit and the environment in the spin-boson model.
\newblock {\em Physical Review A}, 68(3):034301, 2003.

\bibitem{Takahashi1999}
M. Takahashi
\newblock Thermodynamics of one-dimensional solvable models.
\newblock {\em Cambridge University Press}, Cambridge, UK, 1999.

\bibitem{Kondo1964}
J.~Kondo.
\newblock Resistance minimum in dilute magnetic alloys.
\newblock {\em Progress of Theoretical Physics}, 32(1):37--49, 1964.

\bibitem{Leggettetal1987}
A.J. Leggett, S.~Chakravarty, A.T. Dorsey, Matthew~P.A. Fisher, Anupam Garg, and
  W.~Zwerger.
\newblock Dynamics of the dissipative two-state system.
\newblock {\em Reviews of Modern Physics}, 59(1):1--85, 1987.

\bibitem{Haldane1981}
F.D.M. Haldane.
\newblock `{L}uttinger liquid theory' of one-dimensional fluids: {I}. {P}roperties of the {L}uttinger model and their extension to the general {1D} interacting spinless Fermi gas.
\newblock {\em Journal of Physics C: Solid State Physics}, 14:2585--2609, 1981.

\bibitem{vonDelftandSchoeller1998}
J. von Delft and H.~Schoeller.
\newblock Bosonization for beginners - refermionization for experts.
\newblock {\em Annalen der Physik (Leipzig)}, 7(4):225--306, 1998.

\bibitem{CoqblinSchrieffer1969}
B. Coqblin and J.R. Schrieffer
\newblock Exchange interaction in alloys with Cerium impurities.
\newblock {\em Physical Review}, 185(2): 847 ,1969.

\bibitem{CaldeiraandLeggett1983}
A.O. Caldeira and A.J Leggett.
\newblock Quantum Tunnelling in a Dissipative System.
\newblock {\em Annals of Physics}, 149(2):374, 1983.

\bibitem{Baxter1982}
R.J. Baxter.
\newblock {\em Exactly solved models in statistical mechanics}.
\newblock Academic Press Inc., London, 1982.

\bibitem{TsvelickandWiegmann1983}
A.M. Tsvelick and P.B. Wiegmann.
\newblock Exact results in the theory of magnetic alloys.
\newblock {\em Advances in Physics}, 32(4):453--713, 1983.


\bibitem{Jimbo1986}
M. Jimbo.
\newblock Quantum {R}-matrix for the generalized {T}oda system
\newblock {\em Communications in Mathematical Physics}, 102(4):537-547, 1986.

\bibitem{Arnaudonetal2006}
D. Arnaudon, N. Crampe, L. Frappat and E. Ragoucy.
\newblock Spectrum and {B}ethe ansatz equations for the ${U}_q(gl(N))$ closed and open spin chains in any representation
\newblock {\em Ann. H. Poincare}, 7:1217, 2006.

\bibitem{Bazhanov1985}
V.V. Bazhanov.
\newblock Trigonometric solutions of triangle equations and classical {L}ie
  algebras.
\newblock {\em Physics Letters B}, 159B(4,5,6):321--324, 1985.

\bibitem{Afflecketal2001}
I. Affleck and M. Oshikawa and H. Saleur
\newblock Quantum Brownian motion on a triangular lattice and c=2 boundary
   conformal field theory
\newblock {\em Nuclear Physics B}, 594(3):535--606, 2001.

\bibitem{Chamonetal2003}
C. Chamon and M. Oshikawa and I. Affleck
\newblock Junctions of three quantum wires and the dissipative Hofstadter model
\newblock {\em Physical Review Letters}, 91(20):206403, 2003.

\bibitem{Stickneyetal2007}
J.A. Stickney, D.Z. Anderson and A.A. Zozulya
\newblock Transistorlike behaviour of a Bose-Einstein condensate in a triple-well potential
\newblock {\em Physical Review A}, 75:013608, 2007.

\bibitem{Bethe1931}
H.~Bethe.
\newblock Zur {T}heorie der {M}etalle. {E}igenwerte und {E}igenfunktionen der
  linearen {A}tomkette.
\newblock {\em Zeitschrift f\"ur Physik}, 71:205, 1931.

\bibitem{ZarandandvonDelft2000}
G. Zar\'and and J. von Delft.
\newblock Analytical calculation of the finite-size crossover spectrum of the anisotropic two-channel Kondo model
\newblock {\em Physical Review B}, 61(10):6918--6933, 2000.

\bibitem{Amicoetal2008}
Luigi Amico, Rosario Fazio, Andreas Osterloh and Vlatko Vedral.
\newblock Entanglement in many-body systems
\newblock {\em Reviews of Modern Physics}, 80(2):517--576, 2008.

\bibitem{OhandKim2006}
S. Oh and J. Kim.
\newblock Entanglement of an impurity and conduction spins in the Kondo model
\newblock {\em Physical Review B}, 73(5):052407, 2006.


\end{thebibliography}

\begin{appendix}
\setcounter{equation}{0}
\renewcommand{\theequation}{{A}-\arabic{equation}}
\section{Appendix}
\label{sec:appendix}
\subsection{Notation}
Gell-Mann matrices and the standard form of the Lie algebras of $SU(3)$ and $U(3)$ are based on the (multiplicative) algebra of elementary $3\!\times\!3$ matrices, namely $e_{\alpha\beta}e_{\gamma\delta} = \delta_{\beta\gamma} e_{\alpha \delta}$. Thus the commutation relations are
\[
{[}S_{\alpha\beta}, S_{\gamma\delta}{]} = \delta_{\beta\gamma} S_{\alpha \delta} -  \delta_{\alpha \delta} S_{\gamma\beta} .
\]
Introducing the orthogonal basis of trace-normalised $\lambda$-matrices via
$\textstyle{\frac 12}Tr\big(\lambda_{\texttt A}\lambda_{\texttt B}) = \delta_{{\texttt A}{\texttt B}}$,
${\texttt A}, {\texttt B} = 1,2,\cdots, 8$, any $3\!\times\!3$ traceless matrix $x$ can then be expressed in terms of orthogonal coordinates $x_{\texttt{A}}$ via
\[
x_{\alpha\beta} = \textstyle{\frac 12}\sum_{{\texttt A} = {\texttt 1}}^{{\texttt 8}} x_{\texttt A} \big(\lambda_{\texttt A}\big)_{\alpha\beta}, \qquad x_{\texttt A} = \textstyle{\frac 12} Tr(x \lambda_{\texttt A}),
\]
including of course the elementary matrices themselves.
Quantities may also be manipulated using the completeness relation
\[
\delta_{\alpha \beta}\delta_{\gamma\delta} = \textstyle{\frac 13} \delta_{\gamma\beta}\delta_{\alpha\delta} +
 \textstyle{\frac 12}\sum_{{\texttt A} = {\texttt 1}}^{{\texttt 8}}
 \big(\lambda_{\texttt A}\big)_{\gamma\beta}\big(\lambda_{\texttt A}\big)_{\alpha\delta}.
 \]
-- wherein the right hand side may be written uniformly over an extended set of $\lambda$-matrices $\lambda_{{\texttt A}}$, ${\texttt A} = {\texttt 0}, {\texttt 1}, {\texttt 2},\cdots, {\texttt 8}$ by introducing 
$\lambda_{\texttt 0} = \textstyle{\sqrt{\frac{2}{3}}} \, 1_{3\!\times\!3}$ to stand in for the identity matrix. 

Finally in interpreting one-particle operators in second-quantised form, consider the $3$ states $|\alpha\rangle:=c^\dagger_\alpha|0\rangle$ associated with a fixed creation mode (where $\{c^\dagger_\alpha, c_\beta\}=\delta_{\alpha\beta}$ as usual). Then it is easy to check that 
\begin{eqnarray}
\langle\gamma|c^\dagger_\alpha c_\beta|\delta\rangle=\delta_{\delta\alpha}\delta_{\beta\gamma}\equiv (e_{\alpha\beta})_{\delta\gamma},
\end{eqnarray}
-- that is, that the $c^\dagger_\alpha c_\beta$ play the role of elementary matrices on such labelled states. Thus for a term in the particle-impurity interaction such as $\lambda_{\texttt A}\otimes \lambda_{\texttt A}$ we have from above and dropping the $\otimes$,
\begin{eqnarray}
\lambda_{\texttt A}\otimes \lambda_{\texttt A}\rightarrow \sum_{\alpha,\beta}\frac12 Tr (\lambda_{\texttt A} e_{\alpha\beta})c^\dagger_\alpha c_\beta\cdot \lambda_{\texttt A} \equiv \frac12 \sum_{\alpha\beta} c^\dagger_\alpha (\lambda_{\texttt A})_{\alpha\beta}c_{\beta}\cdot \lambda_{\texttt A}.
\end{eqnarray}

\subsection{Charge sector projection}
\label{subsec:Projection}
It was pointed out in the text that the bosonisation transcription was carried out to the neglect of various terms accumulating fermion-number (charge) dependent factors. For example the longitudinal couplings certainly amount to a sum over not only the bosonic modes, which is of course one source of the dissipative coupling, but also contain an explicit number operator term. Similarly the standard expression for the bilinear fermion kinetic energy term (involving as it does a derivative of the fermion field, albeit evaluated at zero) is known to contain a term quadratic in the respective charge operators (in fact the coefficients can also differ for different fermionic boundary conditions, but we do not need this option for our basic derivation).
Overall we assume that the residual fermion number terms amount to an additional contribution from these sources of 
\\[-.3cm]
\begin{align}
\label{eq:QuadraticResidues}
{\texttt C} \sum_\alpha \widehat{N}_\alpha^2 + \sum_\alpha {\texttt C}_\alpha \widehat{N}_\alpha
\equiv & \, 
{\texttt C}\big( \widehat{N}_{{\texttt 3}}^2 + \widehat{N}_{{\texttt 8}}^2\big) + \big({\texttt C}_{{\texttt 3}} \widehat{N}_{{\texttt 3}} + {\texttt C}_{{\texttt 8}} \widehat{N}_{{\texttt 8}}\big) +
\big( {\texttt C} \widehat{N}_{{\texttt 0}}^2 + {\texttt C}_{{\texttt 0}} \widehat{N}_{{\texttt 0}}\big).\nonumber
\end{align}
As mentioned already, total fermion charge is conserved, so for $\widehat{N}_{{\texttt 0}}$ taken fixed at eigenvalue $N_{{\texttt 0}}$ say, the last term is an additive constant. For the remaining terms we turn to the relative conserved quantities $\widehat{N}_{{\texttt 3}} + S_{{\texttt 3}}$,  $\widehat{N}_{{\texttt 8}} + S_{{\texttt 8}}$ and to the projections onto fixed eigenspaces with eigenvalues $M_{{\texttt 3}}$, $M_{{\texttt 8}}$, respectively. Introduce weight labels $| m, y \rangle$ for the basis of the three-dimensional representation of $SU(3)$ corresponding to the the impurity system states, where $m$, $y$ are the eigenvalues of $\lambda_{{\texttt 3}}$, $\lambda_{{\texttt8}}$ (so that $\textstyle{\frac 12}m$, $\textstyle{\frac 12}y$ are the correctly normalised eigenvalues of $\textstyle{\frac 12}\lambda_{{\texttt 3}} = S_{{\texttt 3}}$ and $\textstyle{\frac 12}\lambda_{{\texttt 8}} = S_{{\texttt 8}}$, or isospin and hypercharge, respectively). Imposing the projections, we see that for the total states $| m, y ; \psi \rangle$, the fermionic part $|\psi \rangle$ must have charges
$N_{{\texttt 3}} = M_{{\texttt 3}}-\textstyle{\frac 12}m$, $N_{{\texttt 8}} = M_{{\texttt 8}}-\textstyle{\frac 12}y$ and the charge dependent piece becomes on these states
\begin{align}
{\texttt C}\big( {M}_{{\texttt 3}}^2 + {M}_{{\texttt 8}}^2\big) +  \textstyle{\frac 14}(m^2+y^2)
+ ({\texttt C}_{{\texttt 3}} {M}_{{\texttt 3}} +{\texttt C}_{{\texttt 8}} {M}_{{\texttt 8}}\big) \nonumber\\
- \big({\texttt C} {M}_{{\texttt 3}} + \textstyle{\frac 12}{\texttt C}_{{\texttt 3}}\big) m  
- \big({\texttt C} {M}_{{\texttt 8}} + \textstyle{\frac 12}{\texttt C}_{{\texttt 8}}\big) y  
%
\end{align}
Finally note that the weight basis of the three-dimensional fundamental representation is $|\pm1, 1/\sqrt{3}\rangle$ and $|0, -2/\sqrt{3} \rangle$, so that by construction  $(m^2+y^2) \equiv \textstyle{\frac 43}$ for all states. Thus the first line of the transcription is a further additive constant, while the second line amounts to an external `magnetic' coupling and hence an adjustment to the detuning parameters $\varepsilon_{{\texttt 3}}$, $\varepsilon_{{\texttt 8}}$, by a shift of $-\big({\texttt C} {M}_{{\texttt 3}} + \textstyle{\frac 12}{\texttt C}_{{\texttt 3}}\big)$,
$-\big({\texttt C} {M}_{{\texttt 8}} + \textstyle{\frac 12}{\texttt C}_{{\texttt 8}}\big)$, respectivey.

\subsection{Derivation of ohmic coupling}
\label{subsec:OhmicCoupling}
Given the spectrum of bath frequencies $v_F k = \omega_k \equiv \omega_{n_k} = 2\pi v_F/L \cdot n_k$, from the definition of the spectral frequency $J(\omega)$ we have for any test function $f(\omega)$, from (\ref{eq:Cdefn}),
\begin{align}
\int J(\omega) f(\omega) d\omega = & \, \sum_{n_k} C_{k}^2 f(\omega_k ) 
= 2\pi v_F /L \sum_{n_k} \alpha \omega_{n_k} e^{-\omega_{n_k}/\omega_c} f(\omega_{n_k}) \nonumber \\
\rightarrow & \, 2\pi v_F /L \int dn_k \, \alpha \omega_{n_k} e^{-\omega_{n_k}/\omega_c} f(\omega_{n_k}) 
\equiv \int d\omega \, \big( \alpha \omega e^{-\omega/\omega_c} \big)  f(\omega)
\end{align}
where the approximation that $f(\omega)$ is supported in the region $\omega \ll \omega_c$ has been made. 
The inferred forms of $J(\omega)$ and $\alpha$ are as given in (\ref{eq:JnAlphaDefn}) above. 

\end{appendix}
\end{document}